\documentclass[useAMS,usenatbib]{mn2e}
\usepackage{graphicx,rotate,url,mathptmx,lscape,times}
\def\gtsim{\mathrel{\hbox{\rlap{\hbox{\lower4pt\hbox{$\sim$}}}\hbox{$>$}}}}
\def\lesssim{\mathrel{\hbox{\rlap{\hbox{\lower4pt\hbox{$\sim$}}}\hbox{$<$}}}}

\def\ga{{\rm\thinspace gauss}}

\def\ps{{\rm\thinspace s^{-1}}}
\def\pcc{{\rm\thinspace cm^{-3}}}

\def\Pa{\hbox{{\rm P}$\alpha$}}

\def\htwo{\hbox{{\rm H}$_2$}}
\def\hplus{\hbox{{\rm H}$^+$}}
\def\h0{\hbox{{\rm H}$^0$}}
\DeclareMathAlphabet{\vib}{OML}{cmm}{m}{it}

\title[\htwo\ emission in cooling-flow filaments]{The origin of molecular hydrogen emission in cooling-flow filaments\thanks{Contains material \copyright\ British Crown copyright 2007/MoD}}

\author[G.J. Ferland, et al.]
       {\parbox[]{6.0in}
        {G.J. Ferland$^{1,2}$\thanks{E-mail: gary@pa.uky.edu},
        A.C. Fabian$^1$,
        N.A. Hatch$^{3}$,
        R.M. Johnstone$^1$,\\
        R.L. Porter$^{1,2}$,
        P. A. M. van Hoof$^4$,
        R.J.R. Williams$^{5}$\\
        \footnotesize
        $^1$Institute of Astronomy, University of Cambridge, Madingley
        Road, Cambridge CB3 0HA\\
        $^2$Department of Physics, University of Kentucky, Lexington, KY 40506, USA\\
        $^3$Leiden Observatory, University of Leiden, P.B. 9513, Leiden 2300 RA, The Netherlands\\
        $^4$Royal Observatory of Belgium, Ringlaan 3, 1180 Brussels, Belgium\\
        $^5$AWE plc, Aldermaston, Reading RG7 4PR}}

\date{
      Received }

\pagerange{\pageref{firstpage}--\pageref{lastpage}}
\pubyear{}

\begin{document}

\maketitle

\label{firstpage}

\begin{abstract}

\noindent
The optical filaments found in many cooling flows in galaxy clusters consist of low density ($\sim 10^3 \pcc$) cool ($\sim 10^3$ K) gas
surrounded by significant amounts of cosmic-ray and magnetic-field energy.
Their spectra show anomalously strong low-ionization and molecular emission lines when
compared with galactic molecular clouds exposed to ionizing radiation such as
the Orion complex.
Previous studies have shown that the spectra cannot be produced by O-star photoionization.
Here we calculate the physical conditions in dusty gas
that is well shielded from external sources of ionizing photons and is
energized either by cosmic rays or dissipative MHD waves.
Strong molecular hydrogen lines, with relative intensities similar to those observed, are produced.
Selection effects introduced by the microphysics produce a correlation between the
\htwo\ line upper level energy and the population temperature.
These selection effects allow a purely collisional gas to produce \htwo\ emission
that masquerades as starlight-pumped \htwo\ but with intensities that are far stronger.
This physics may find application to any environment where a broad range of gas densities or
heating rates occur.
\end{abstract}


\section{Introduction}
\label{intro}

The origin of the optical filaments often found near central regions of massive clusters of galaxies
remains mysterious (see \citealt{JohnstoneEtAl07} for a discussion of the extensive literature).
The presence of dust and blue light,
particularly in the centers of these objects,
suggests that star formation has occurred, at least in some regions.
Their spectra are quite different from H~II regions,
which we take as representative of emission from gas near newly-formed stars.
Low-ionization optical forbidden lines and
near-IR H$_2$ lines are very strong relative to hydrogen recombination lines.
No one model can account for these observations \citep{JohnstoneEtAl07}.

The combination of Spitzer and ground-based observations allows H$_2$ lines from a wide range of
excitation energies to be studied (\citealt{JohnstoneEtAl07}).
The resulting \htwo\ population-excitation diagram shows that
the level population temperatures correlate with upper-level energies.
The high-excitation lines indicate $T_{\rm{pop}} \approx 2000$~K, substantially hotter than
lines from low-lying levels, with $T_{\rm{pop}} \approx 300$~K.
A similar effect occurs in PDRs near galactic H~II regions
due to starlight photoexcitation of higher levels (\citealt{vanDishoeck04}).
The lines produced by this process are far weaker than is seen in these filaments.
Generally, radiation from early-type stars cannot produce extremely strong \htwo\ lines because the
luminosity in the lines is, at most,
a small fraction of the stellar luminosity in the 1200 -- 912 \AA\ range
\citep{Sternberg05}.
Other energy sources are needed.
Shocks due to stellar outflows are one possibility
\citep{FernandesEtAl07,JaffeEtAl01,WilmanEtAl02}.

At least two sources of energy exist that are unique to the cooling-flow environment.
The filaments are surrounded by a mix of hot thermal particles and synchrotron-emitting
relativistic particles \citep{SandersFabian07}.
The latter are referred to as cosmic rays here.
Additionally, Faraday rotation measures and the long linear geometries suggest that magnetic
fields are important \citep{GuidettiEtAl07,TaylorEtAl07}.

Here we investigate whether non-radiative heating,
produced by cosmic rays or dissipative MHD waves, could produce the observed emission.
Both would deposit energy into otherwise well-shielded molecular gas.
We consider the cores of the filaments, ignoring complications such
as a possible ionized sheath that would be produced by \emph{in situ} star formation
or mixing layers \citep{CrawfordFabian92}.
We show that gas energized by non-radiative heating processes can
produce an \htwo\ spectrum that is very similar to those that are observed.

\section{Conditions in shielded cores}

Table \ref{tab:H2SpectraOrionFlow} compares \htwo\ line intensities
relative to \Pa\
in the Perseus cooling flow with those in the Orion complex.
Intensities for the filaments are taken from Table 8 of
\citep{JohnstoneEtAl07} while those for
the Orion Bar are from \citet{SellgrenEtal90} and \citet{AllersEtal05}.
The Orion Bar is an \hplus / \h0 / \htwo\ interface viewed roughly edge on.
Its successive layers of ionization are displaced on the sky,
as shown in Section 8.5 of Osterbrock \& Ferland (2006, hereafter AGN3).
The separate regions are not spatially resolved in the extragalactic case, so
the intensities across the Bar were co-added to create the values listed in
Table \ref{tab:H2SpectraOrionFlow}.
This procedure is not more accurate than a factor of two.
Even so, the differences in the line intensities are dramatic.
The \htwo\ lines are far stronger in the filaments than is found in
this prototypical H~II region.
This suggests that additional energy sources are active in the cooling flow filaments.

\begin{table}
\centering
\caption{\htwo\ line intensities in a filament in the NGC1275 cooling-flow and the Orion Bar}
\begin{tabular}{lllllllllll}
\hline

\htwo\ Line & I / I(P$\alpha$) flow &  I / I(P$\alpha$) Orion \\
\hline
0-0\,S(1) 17.03 \micron & 0.65 & 1$\times 10^{-2}$ \\
0-0\,S(2) 12.28 \micron & 0.30 & 9$\times 10^{-3}$ \\
1-0\,S(1) 2.121 \micron & 0.70 &  6$\times 10^{-3}$\\

\hline
\end{tabular}
\label{tab:H2SpectraOrionFlow}
\end{table}

Could cosmic rays or extra heating, perhaps due to dissipative MHD waves, produce the
strong molecular emission?
The environment inside a real filament is likely to have a mix of particle densities and
be exposed to a range of
cosmic rays, dissipative MHD waves, and possibly starlight.
We consider only the cosmic ray and wave heating cases here.
We do not address questions such as the transport
of cosmic rays into possibly magnetically confined filaments,
or the process by which kinetic energy in MHD waves might be converted into thermal energy.
Rather we follow the effects of these energy sources on the microphysics of
the gas and predict the resulting \htwo\ line intensities.

We use the development version of the spectral synthesis code
Cloudy, last described by \citet{Ferland98}.
AGN3 discuss much of the physics of the environment with extensive reviews of
plasma simulation codes.
Our treatment of the H$_2$ molecule, the focus of this paper,
is described in \citet{ShawEtal05}.

Our treatment of cosmic rays is described in \citet{FerlandMushotzkyCR84},
\citet{AbelEtal05}, \citet{ShawEtal06} and \citet{ShawEtal07}.
In molecular gas a primary cosmic ray will produce a shower of secondary
electrons which ionize and excite atoms and molecules.
Relatively little of the
energy goes into heating the gas (AGN3).
Conversely, if the gas is highly ionized then nearly all the cosmic-rays energy will be
thermalized and little ionization or excitation occurs.

We also consider an ``extra heating'' case where thermal energy,
perhaps produced by dissipative MHD waves, is injected.
This energy, ultimately due to the kinetic energy of a filament, is added to the thermalized kinetic energy of the particles in the material.  We assume that the combination of plasma stream instabilities and particle-particle collisional coupling is effective enough in maintaining a Maxwellian distribution for the particle distribution functions that the detail of the energy injection process is unimportant.
Ion-neutral drift, where charged particles are coupled to the field and move relative to
neutral particles, might be one source of such heat
\citep{LoewensteinFabian90}.
The effects are quite different from cosmic rays since the
heating has no corresponding direct ionization or internal excitation.
Ionization and excitation only occurs when the kinetic
temperature becomes high enough to overcome energetic thresholds.
Comparing the two cases, at a given kinetic temperature gas energized by cosmic rays will be more
highly ionized and excited than the ``extra-heating'' case.

We parameterize the cosmic ray case by the cosmic-ray density
relative to the galactic background value.  Background cosmic rays produce
an H$^0$ ionization rate of $2.5 \times 10^{-17}$ s$^{-1}$ in gas
with a low electron fraction \citep{WilliamsElAl98}.
The range of relative cosmic ray rates includes
the values measured by
\citet{SandersFabian07} and is summarized in Section 3 below.
In the ``extra heating'' case, energy is added to the thermal kinetic energy of the particles at a rate specified in $\rm erg\,cm^{-3}\,s^{-1}$.
It is not now possible to unambiguously convert a measured magnetic field into a
wave dissipation rate \citep{HeilesCrutcher05} so we only show the
range of rates with observable effects.

H$_2$ emission is likely to be produced in shielded cores where any ionizing radiation
produced by either star formation or cooling-flow emission has been extinguished by
surrounding gas.
The $z = 0$ metagalactic background is the radiation field incident upon the cloud.
This includes the CMB and the 2005 version of the \citet{HaardtMadau96} background
with both starburst and quasar continua.
This continuum is attenuated by a dusty gas with a total hydrogen column density of 10$^{21}$ cm$^{-2}$.
The extinguished continuum has little effect on the simulations presented below
but is included for completeness.
For the ``extra-heating'' case we also include cosmic rays with the galactic background value.

There is no direct
measurement of the gas metallicity or dust to gas ratio in the filaments.
For simplicity we assume galactic interstellar medium gas-phase abundances and dust
\citep{AbelEtal05}.
Many of the results presented below do not depend critically on these assumptions.
Molecular hydrogen forms on grain surfaces in dusty environments.
The rate of formation depends on the grain size distribution, temperature,
composition and dust-to-gas ratio,
all of which are unknown.
We adopt
the galactic ISM catalysis rate of
$3 \times 10^{-17}$ cm$^3$ s$^{-1}$ \citep{Jura75}.

We are interested in molecular regions that must be well shielded from ionizing radiation.
The absorbing column imposed on the metagalactic radiation field ensures that little light
shortward of 912\AA\ penetrates.
We further assume ``Case~B'' (AGN3),
that resonance lines of H~I and H$_2$ have large optical depths because of
this large column density.
The result is the Lyman lines do not escape and continuum fluorescent pumping of H I and \htwo\ resonance lines is not important.
Absorption of UV radiation by electronic transitions is the main \htwo\ destruction process in
star-forming regions but will be unimportant in the calculations presented below
because of the shielding provided by the large column density.

With these assumptions the free parameters are the hydrogen density $n_{\rm H}$ (cm$^{-3}$)
and the cosmic ray or extra heating rates.
Pairs of figures, giving the cosmic ray case as the upper panel and the
``extra-heating'' case as the
lower one, will be shown below.
The hydrogen density is the x-axis in all cases.

\begin{figure}
\protect\resizebox{\columnwidth}{!}
{\includegraphics{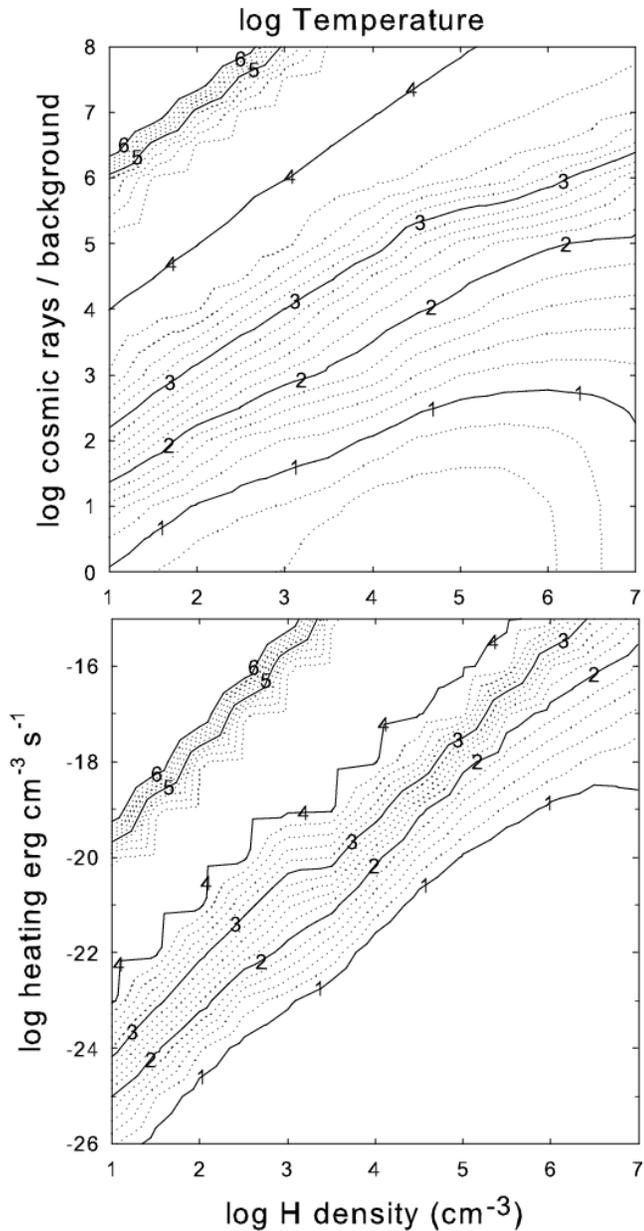}}
\caption{The log of the computed gas kinetic temperature is shown as a function of the hydrogen
density and (top) the cosmic ray rate relative to the galactic background
and (bottom) the extra heating rate.}
\label{fig:temperature}
\end{figure}

Figure \ref{fig:temperature} shows the log of the gas kinetic temperature as a function of
these parameters.
The temperature is the result of heating and cooling processes.
In the cosmic ray case the heating per unit volume is proportional to
$r_{\rm{CR}} n_{\rm H}$, where $r_{\rm{CR}}$ is the primary cosmic ray ionization rate and $n_{\rm H}$
is the hydrogen density.
The cooling per unit volume is proportional to the collision rate, or $ n_{\rm H}^2$.
The kinetic temperature, usually increasing with the ratio of heating to cooling,
is then proportional to $r_{\rm{CR}} / n_{\rm H}$.
Lines of constant temperature tend to run at a roughly 45 degree angle corresponding
to a constant $r_{\rm{CR}} / n_{\rm H}$ ratio.
The curves turn over at the highest densities when collisional cooling is suppressed.

In the ``extra-heating'' case the heating per unit volume, $G_{\rm{extra}}$, is
assumed to be independent of density for simplicity.
The temperature-determining ratio of heating to cooling is then proportional to
$G_{\rm{extra}} / n_{\rm H}^2 $.
Contours of constant temperature run at a steeper angle as a result.

In both cases the temperatures in the lower right hand corner of Figure \ref{fig:temperature}
are close to the CMB.
Temperature increases as the heating increases,
reaching the highest values in the upper left corner where the gas is hot and highly ionized.

\begin{figure}
\protect\resizebox{\columnwidth}{!}
{\includegraphics{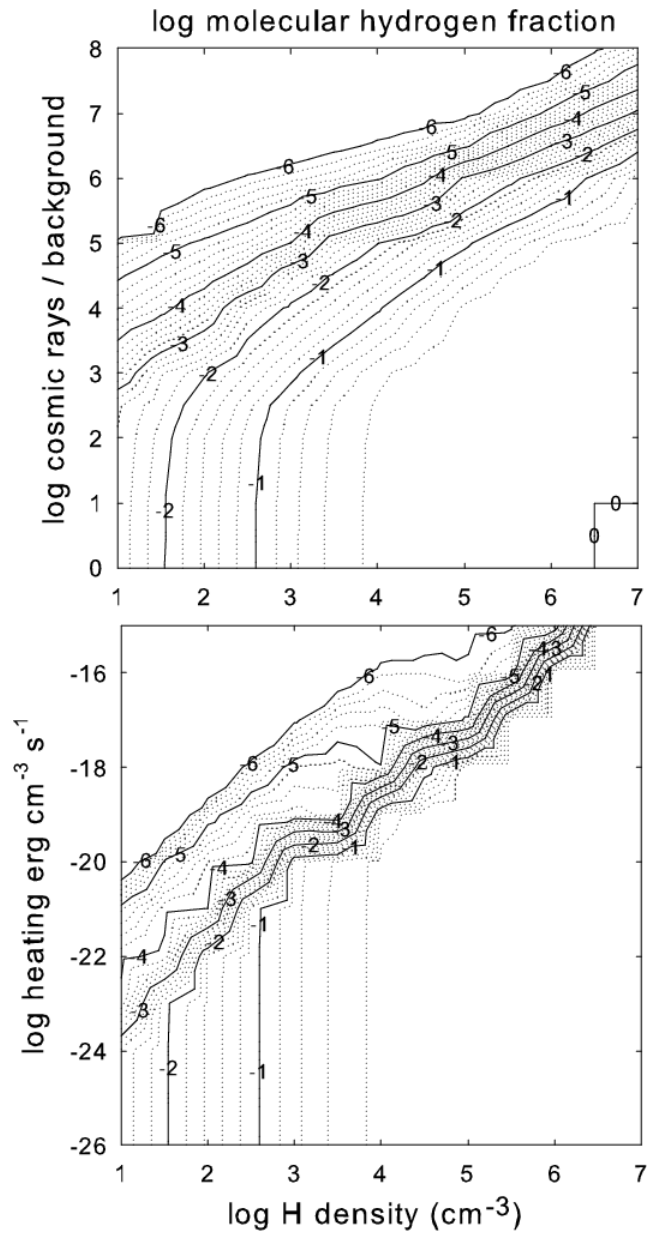}}
\caption{The log of the hydrogen molecular fraction for the cosmic ray (top)
and ``extra-heating'' (bottom) cases.}
\label{fig:HMoleFrac}
\end{figure}

Figure \ref{fig:HMoleFrac} shows the log of the hydrogen molecular fraction.
The lower right corner is predominantly molecular.
Gas becomes increasingly dissociated and eventually ionized along a diagonal extending to
the upper left.
In the cosmic ray case \htwo\ predominantly dissociates following
suprathermal excitation to molecular triplet states.
Excitations to the singlets dissociate roughly 10\% of the time with
the remainder decaying into various levels within the ground electronic state
\citep{SternbergEtAl87,DalgarnoEtAl99}.
Thus cosmic rays dissociate and excite \htwo\ even when the kinetic temperature is low.
This is in contrast with the ``extra-heating'' case,
where little excitation or dissociation is produced until
the kinetic temperature becomes large enough for thermal collisions to drive the processes.

\begin{figure}
\protect\resizebox{\columnwidth}{!}
{\includegraphics{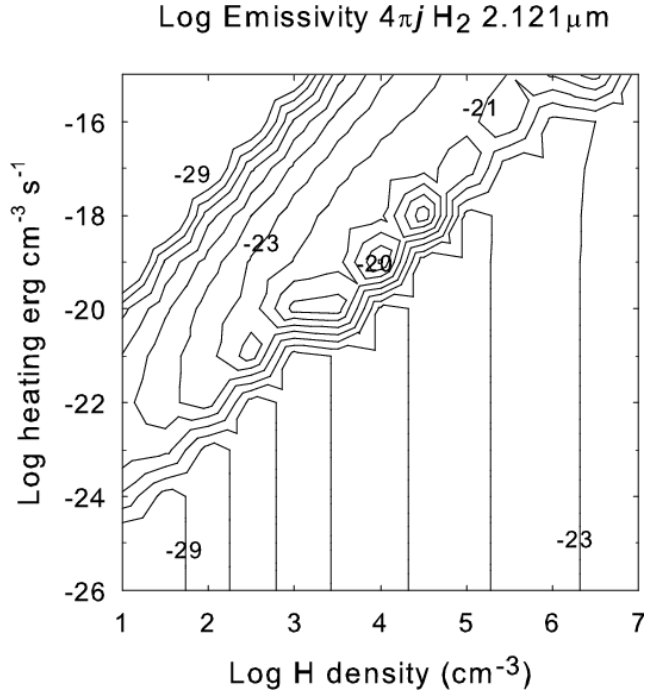}}
\caption{The emissivity of the H$_2$ 2.121\micron line
for the ``extra-heating'' case.  Contours are drawn at 1 dex intervals,
ranging from a minimum of $log 4 \pi j$ of -29 up to -20. }
\label{fig:emissivity}
\end{figure}

Figure \ref{fig:emissivity} shows the log of the emissivity $4 \pi j$
(erg cm$^{-3}$ s$^{-1}$) of the 1-0 S(1) 2.121 $\micron$ line,
one of the strongest \htwo\ lines in the NIR spectrum.
The ``extra-heating'' case is shown to illustrate the effects of changing temperature.
The emissivity is strongly peaked towards a narrow band running along lines of roughly constant
temperature and H$_2$ fraction.
Neglecting second-order effects, the emissivity of the line,
which is predicted to be collisionally excited, will be proportional to
the product of the density  $n$(H$_2$) and a Boltzmann factor

\begin{equation}
4\pi j \propto n\left( {{\rm{H}}_2 } \right)\exp \left( { - \chi /T} \right)
\label{equation:t_ex}
\end{equation}
where $\chi \approx 7000$~K is the excitation energy of
the upper level of the transition.
The gas is mostly molecular in the lower right corner of the diagram but the temperature is too
low to excite the transition.
The temperature is high in the upper left corner but there is little H$_2$.
The line is emitted efficiently across a narrow band where the product of the H$_2$ density and
Boltzmann factor is large.
The ridge of peak emissivity occurs at a gas kinetic temperature of roughly 2000~K.

This result can be generalized.
Because of the form of equation \ref{equation:t_ex}, a Boltzmann factor
that increases exponentially as temperature rises,
and an \htwo\ density that decreases as the temperature increases,
the emissivity of a collisionally-excited line will peak when
the kinetic temperature is roughly half of the excitation energy.
The large changes in emissivity
shown in Figure \ref{fig:emissivity} suggest that,
if gas exists across the parameter space,
a linear detector like a spectrometer will pick out only the peaks.
Similar effects are present in emission lines of active galactic nuclei
\citep{BaldwinEtAl95}.

\begin{figure}
\protect\resizebox{\columnwidth}{!}
{\includegraphics{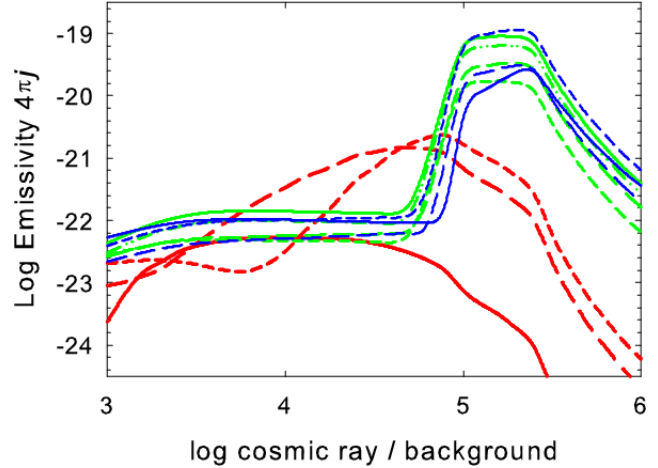}}
\caption{The predicted emissivities of all \htwo\ lines observed by \citet{JohnstoneEtAl07}
are shown as a function of the cosmic-ray density.
The density is $n_{\rm H} = 10^4$ cm$^{-3}$.
The low-excitation rotational lines are (red)
28.21 \micron\ (solid), 17.03 \micron\ (long dash), and 12.28 \micron\ (short dash).
The mid-excitation (2000 K $< E_U <$ 8000 K) lines are (green)
2.121 \micron\ (solid), 2.033 \micron\ (long dash), 2.223 \micron\ (short dash),
and 2.423 \micron\ (dash dot).
High excitation ($E_U > 8000$ K) lines are (blue) 1.748 \micron\ (solid), 1.891 \micron\ (long dash),
and 1.957  \micron\ (short dash).
The emissivity of each line peaks at a temperature that is proportional to its upper level energy, as found by \citet{JohnstoneEtAl07}.
}
\label{fig:varCRden4}
\end{figure}

We now concentrate on the cosmic-ray case since their densities have been
measured in regions of the flow \citep{SandersFabian07}.
Figure \ref{fig:varCRden4} shows the emissivities,
the emission per unit volume (erg $\pcc \ps $ ),
of all \htwo\ lines detected in \citet{JohnstoneEtAl07}.
These would be multiplied by the cloud volume, or equivalently the mass of \htwo , to obtain line luminosities.
We adopt the \citet{JohnstoneEtAl07} density of
$n_{\rm H} = 10^4$ cm$^{-3}$ and vary the cosmic-ray density.
This could occur if gas were trapped in magnetic field lines, so that $n_{\rm H}$ is constant,
but exposed to a range of cosmic ray fluxes, perhaps due to attenuation or shielding.
This corresponds to a vertical slice through Figures 1 -- 3.

\citet[Table 11]{JohnstoneEtAl07} demonstrate a correlation between a level's energy and its population temperature $T_{\rm{pop}}$,
the temperature corresponding to the population distribution.
If the levels are thermalized then $T_{\rm{pop}}$ is the kinetic temperature.
Figure \ref{fig:varCRden4} shows the predicted emissivities of
all lines reported in that paper.
If the levels are thermalized then the population temperature will
be weighted towards $T_{\rm{peak}}$, the
kinetic temperature where the emissivity peaks.

The distribution of emissivities shown in Figure \ref{fig:varCRden4} reproduces the
\citet{JohnstoneEtAl07} correlation.
The observed ($T_{\rm{pop}}$) and predicted ($T_{\rm{peak}}$) peak temperatures are,
for low-excitation levels, $T_{\rm{pop}} = 300 \pm 20 $~K, $T_{\rm{peak}} = 610 \pm 450$~K,
for intermediate excitation, $T_{\rm{pop}} = 1730 \pm 250$~K, $T_{\rm{peak}} = 2300 \pm 400$~K,
and for high excitation, $T_{\rm{pop}} = 2580 \pm 150$~K, $T_{\rm{peak}} = 2900 \pm 400$~K.
This agreement, generally within the scatter, is partially fortuitous because the
predicted kinetic temperature
depends on the gas metallicity, which we simply set to galactic ISM abundances and depletions.
The overall trend, where population temperature correlates with excitation temperature, is inescapable
for gas with a non-radiative energy source.

\section{Conclusions}
\label{Discussion and conclusions}

The comparison between the relative intensities of \htwo\ and H~I lines in the Orion star-forming region and the cooling-flow filament studied by \citet{JohnstoneEtAl07}
(Table \ref{tab:H2SpectraOrionFlow})
shows that the \htwo\ lines are far stronger in the filament.
Starlight fluorescence produces only weak \htwo\ emission since only a small portion of the
Balmer continuum can be converted into \htwo\ lines.
This strongly suggests that the lines are formed by more efficient, perhaps non-radiative,
processes in cooling flows.

Both cosmic rays and dissipative MHD waves are candidates as extra energy sources.
Powerful selection effects introduced by the microphysics cause each emission line
to have a peak emissivity at a temperature related to its upper level energy.
Both the great strength of the \htwo\ lines and the observed
correlation between excitation potential and population temperature result.
Actually, any heating source that simultaneously heats and dissociates \htwo\ would
produce similar effects.

The selection effects cause a collisionally-excited \htwo\ spectrum to masquerade
as classical starlight-induced fluorescence.
In the radiative case the lowest-energy levels are collisionally populated and their populations
indicate the kinetic temperature.
Higher levels are fluorescence excited and have a physically meaningless higher temperature
\citep{vanDishoeck04}.
A similar range of population temperatures is produced in the non-radiative cases considered here,
but all of the temperatures are real.

The observed cosmic ray densities are large enough to produce the observed emission
if $n_{\rm H}$ is lower than usually assumed.
\citet{SandersFabian07} find a cosmic ray energy density of roughly 150 eV cm$^{-3}$ for
central regions of the Perseus cluster.
For comparison \citet{Webber98} finds the density of relativistic electrons
in the local interstellar medium to be 0.2 eV cm$^{-3}$.
The ratio of cluster to local ISM CR ionization rates, the vertical axis in most of the plots
in this paper, is of order 10$^3$.
The predictions scale roughly as the ratio $r_{\rm{CR}} / n_{\rm H}$.
The peak emissivity for high-excitation lines (Figure \ref{fig:varCRden4}),
occurs at $r_{\rm{CR}} \approx 10^{5.5}$ for a density of $n_{\rm H} = 10^4$ cm$^{-3}$.
A density of $n_{\rm H} = 10$ cm$^{-3}$ would have similar spectra at $r_{\rm{CR}} \approx 10^3$.
Since \citet{JohnstoneEtAl07} was published, new calculations show that some
\htwo\ collision rates were underestimated by roughly 2 dex \citep{WrathmallEtAl07}.
Simple scaling relations suggest that
this would lower $n_{\rm H}$ and the gas pressure by a similar amount.
This would solve the long-standing puzzle posed by the high gas pressures associated with
the high density \citep{JohnstoneEtAl07}.
The filaments' gas pressure would then be in line with the pressure
in the surrounding hot plasma.

The emissivities produced by the non-radiative processes we discuss are similar to those
assumed by \citet{JohnstoneEtAl07}.  Hence the mass of \htwo\ required to account for the
observed emission will be similar.
The total mass in warm \htwo , roughly $10^5\ M_{\sun} $ \citep{JohnstoneEtAl07},
is a tiny fraction of the mass of cold gas, $\sim 4 \times 10^{10}\ M_{\sun} $,
reported by \citet{SalomeEtAl06}.
This suggests that the warm \htwo\ may be produced in small regions which are
affected by the localized deposition of energy in the forms of
dissipative MHD waves, cosmic rays, or shocks.
We have shown here that such sources of energy can account for the observed spectra.

\section{Acknowledgments}
ACF acknowledges support by the Royal Society.  RMJ
acknowledges support by the Royal Society and STFC.
GJF thanks the NSF (AST 0607028), NASA (NNG05GD81G), STScI
(HST-AR-10653) and the Spitzer Science Center (20343) for support.
PvH acknowledges support from the Belgian Science Policy Office (grant MO/33/017).
NAH acknowledges funding from the Royal Netherlands Academy of Arts
and Sciences.
We thank the referee for a helpful review of the paper.

\bsp

\label{lastpage}
\clearpage
\end{document}